\documentstyle[preprint,aps,12pt]{revtex}
\tightenlines
\input epsf
\begin{document}
%\draft command makes pacs numbers print
\draft

%\preprint{YITP-98-37,\hep-ph/9807378}
\preprint{\vtop{{\hbox{YITP-01-9}%\vskip-0pt\hbox{hep-ph/01?????}
}}}

\thispagestyle{empty}

\title{On the $D^+\rightarrow\omega\pi^+$ decay}
 
\author{ K. Terasaki}
\address{Yukawa Institute for Theoretical Physics, 
Kyoto University, Kyoto 606-8502, Japan}
%\date{January 2001}
\maketitle

\begin{abstract} % insert abstract here
The $D^+\rightarrow\omega\pi^+$ decay is studied by decomposing its 
amplitude into a sum of factorizable and non-factorizable ones. 
Its rate is predicted as a function of mass and width of a
hypothetical hybrid meson. 
\end{abstract}
 
\vskip 0.5cm
%\narrowtext
% body of paper here
%%%%%%%%%%%%%%%%%%%%%%%%%%%%%%%%%%%%%%%%%%%%%%%%%%%%%%%%%%%%
\newpage

%\section{Introduction}

(Quasi) two body decays of charm mesons can be described in terms of 
three different quark-line diagrams, i.e., spectator, color 
mismatched spectator and annihilation diagrams in the weak boson mass 
$m_W\rightarrow\infty$ limit. As is well known, the naively 
factorized amplitudes given by the last two diagrams are 
suppressed (color suppressed and helicity suppressed, respectively) 
in the BSW scheme~\cite{BSW} while the measured decay rates for 
color suppressed $D^0\rightarrow\bar K^0\pi^0$, $\bar K^{*0}\pi^0$, 
etc. are not very small against the expectation. Therefore, 
the factorization has been implemented by taking into account final 
state interactions. However, amplitudes with final state interactions 
are given by non-leading terms in the large $N_c$ expansion which are 
not factorizable~\cite{BCP-3,Dalitz,Nova}. It implies that the 
non-factorizable contributions play an essential role, at least,
in hadronic weak decays of charm mesons. 

In this short note, we study the $D^+\rightarrow\omega\pi^+$ decay 
decomposing its amplitude into a sum of factorizable and 
non-factorizable ones as in two body decays~\cite{TBD99}. Therefore, 
our starting point is to decompose the effective weak Hamiltonian 
into a sum of the BSW Hamiltonian, $H_w^{\rm BSW}$, and an extra 
term, $\tilde H_w$, i.e., 
$H_w \rightarrow H_w^{\rm BSW} + \tilde H_w$, where $H_w^{\rm BSW}$ 
and $\tilde H_w$ are responsible for factorizable and 
non-factorizable amplitudes, respectively. 
For more details, see Refs.~\cite{BCP-3,Dalitz,Nova,TBD99}. 

The amplitude for the $D^+\rightarrow\omega\pi^+$ decay includes all 
the three types of amplitudes mentioned above. Using the naive 
factorization in the BSW scheme, we calculate factorizable amplitudes 
$M(D\rightarrow VP)_{\rm FA}$ for $D\rightarrow VP$ decays, 
where $D$, $V$ and $P$ denote a charm, a vector and a pseudo-scalar 
(PS) meson, respectively. A typical result is given in Table~I, in 
which ${\cal A}$ is defined by, for example, 
%%%%%%%%%%%%%%%%%%%%%%%%%%%%%%%%%%%%%%%%%%%%%%%%%%%%%%%%%%%%%%%%%%%%
$M(D^+\rightarrow \omega\pi^+)_{\rm FA}
=-i({G_F/ \sqrt{2}})V_{cd}V_{ud}a_1f_\pi 
{\cal A}(D^+\rightarrow \omega\pi^+)_{\rm FA}$, etc. 
%%%%%%%%%%%%%%%%%%%%%%%%%%%%%%%%%%%%%%%%%%%%%%%%%%%%%%%%%%%%%%%%%%%%
The CKM matrix elements~\cite{CKM} are taken to be real in this note. 
In the above, matrix elements of currents have been taken in the same 
form as in Ref.~\cite{Nova}. In both of the annihilation and the 
color suppressed decays, the factorized amplitudes are actually 
suppressed. 

Next, we study non-factorizable amplitudes. As discussed in 
Refs.~\cite{Dalitz} and \cite{Nova}, they are dominated by dynamical 
contributions of various hadron states and can be estimated by using 
a hard PS-meson approximation in the infinite momentum frame 
(IMF)~\cite{hardpion,suppl}. In this approximation, the 
non-factorizable amplitude for $D \rightarrow V\bar P$ is given  by 
$M \simeq M_{\rm ETC} + M_{\rm S}$, 
where $M_{\rm ETC}$ is written in the form, 
%%%%%%%%%%%%%%%%%%%%%%%%%%%%%%%%%%%%%%%%%%%%%%%%%%%%%%%%%%%%%%%%%%%%%
$M_{\rm ETC}(D\rightarrow V\bar P)
= ({i/f_{P}})\langle{V|[V_{P}, \tilde H_w]|D}\rangle$, 
%%%%%%%%%%%%%%%%%%%%%%%%%%%%%%%%%%%%%%%%%%%%%%%%%%%%%%%%%%%%%%%%%%%%%
by using $[V_P + A_P, \tilde H_w]=0$. $V_P$ is an $SU_f(3)$ 
charge and $A_P$ is its axial counterpart. $M_{\rm S}$ is given by a 
sum of all possible pole amplitudes, 
%%%%%%%%%%%%%%%%%%%%%%%%%%%%%%%%%%%%%%%%%%%%%%%%%%%%%%%%%%%%%%%%%%%%%%
$M_{\rm S} = \sum_n M^{(n)} + \sum_l M^{(l)}$, 
%%%%%%%%%%%%%%%%%%%%%%%%%%%%%%%%%%%%%%%%%%%%%%%%%%%%%%%%%%%%%%%%%%%%%
where 
%%%%%%%%%%%%%%%%%%%%%%%%%%%%%%%%%%%%%%%%%%%%%%%%%%%%%%%%%%%%%%%%%%%%%%  
\begin{eqnarray} 
&&M^{(n)}(D\rightarrow  V\bar P)
=-{i \over f_{P}}\Bigl({m_{V}^2 - m_{D}^2 
                                 \over m_n^2 - m_{D}^2}\Bigr)
  \langle{V|A_{P}|n}\rangle
                   \langle{n|\tilde H_w|D}\rangle ,  \nonumber\\
&&M^{(\,l\,)}(D\rightarrow  V\bar P)
=-{i \over f_{P}}\Bigl({m_{V}^2 - m_{D}^2 
                  \over m_l^2 - m_{V}^2}\Bigr)
\langle{V|\tilde H_w|\,l\,}\rangle
            \langle{\,l\,|A_{P}|D}\rangle . 
                                                    \label{eq:SURF}
\end{eqnarray}
%%%%%%%%%%%%%%%%%%%%%%%%%%%%%%%%%%%%%%%%%%%%%%%%%%%%%%%%%%%%%%%%%%%%
The intermediate $n$ and $l$ run over all possible single meson
states but the states $\langle n|$ and $|l\rangle$ which sandwich 
$\tilde H_w$ with $|D\rangle$ and $\langle V|$, respectively, should 
conserve spins~\cite{Pakvasa}. For the same reason, we discard 
$M_{\rm ETC}(D\rightarrow V\bar P)$ which is proportional to 
$\langle{V|\tilde H_w|D}\rangle$. 

Four-quark $\{qq\bar q\bar q\}$ mesons in the $s$- and $u$-channels 
of the spectator decays, $\{qq\bar q\bar q\}$ and ordinary 
$\{q\bar q\}$ (and hybrid $\{q\bar qg\}$) mesons in the $s$- and 
$u$-channels, respectively, of the color mismatched spectator decays, 
and $\{q\bar q\}$ (and hybrid $\{q\bar qg\}$) and 
$\{qq\bar q\bar q\}$ mesons in the $s$- and $u$-channels,
respectively, of the annihilation decays can contribute. 
However, it has been known that the factorized amplitude 
dominates~\cite{BSW} the one for the $D_s^+\rightarrow\phi\pi^+$, 
which 
%%%%%%%%%%%%%%%%%%%%%%%%%%%%%%%%%%%%%%%%%%%%%%%%%%%%%%%%%%%%%%%%%%%%
\newpage
\begin{center}
\begin{quote}
{Table~I. Factorized and non-factorizable amplitudes for 
typical quasi two-body decays, where 
$h_D=-\sqrt{2}\langle{D^{*0}|A_{\pi^-}|D^+}\rangle$ 
and $\hat g_H=\langle{\omega|A_{\pi^-}|\hat\pi_H^+}\rangle /\sqrt{2}$.  
The amplitudes ${\cal A}_{\rm FA}$ and ${\cal M}_{\rm NF}$ are 
defined in the text.}
\end{quote}
\vspace{0.5cm}

\begin{tabular}
{c|c|c}
\hline
 & \vspace{-3.5mm}\\
$\quad\,\,${\rm Decay}
& {$ {\cal A}_{\rm FA}\,$}
& {$ {\cal M}_{\rm NF}\,$}
\\
&\vspace{-3.5mm}&\\
\hline 
 & \vspace{-3.5mm}&\\
$D_s^+\rightarrow\phi\pi^+$
& $\displaystyle{
A_0^{(\phi D_s)}(m_\pi^2) 
2m_{\phi}[\epsilon^{(\phi)*}(p')\cdot p]
}$
&  neglected
\\
 & \vspace{-3.5mm}&\\
\hline
 & \vspace{-3.5mm}&\\
$D_s^+\rightarrow\rho^0\pi^+$
& $0$
& $0$
\\
 & \vspace{-3.5mm}&\\
\hline
 & \vspace{-3.5mm}&\\
$D_s^+\rightarrow\omega\pi^+$
&  small 
& $\displaystyle{
-{\langle{\hat\pi_H^+|\tilde H_w|D_s^+}\rangle \over f_\pi}
\Biggl({m_{D_s}^2 - m_\omega^2 
\over m_{D_s}^2 - m_{\hat\pi_H}^2}\Biggr)\sqrt{2}\hat g_H
}$
\vspace{-3.5mm}
\\
 & &\\
\hline
 & \vspace{-3.5mm}&\\
$D^+\rightarrow\phi\pi^+$
& small
&
{$\displaystyle{
-{\langle{\,\phi\,\,|\tilde H_w|D^{*0}}\rangle \over f_\pi}
\Biggl({m_{D}^2 - m_\phi^2 
\over m_{D^*}^2 - m_{\phi}^2}\Biggr)\sqrt{1\over 2}h_D}
$}
\vspace{-3.5mm}
\\
 & &\\
\hline
 & \vspace{-3.5mm}&\\
$D^+\rightarrow\omega\pi^+$
& $\displaystyle{
\sqrt{1\over 2}A_0^{(\omega D)}(m_\pi^2) 
2m_{\omega}[\epsilon^{(\omega)*}(p')\cdot p]
}$ 
& \hspace{0mm} 
\begin{tabular}{l}
{$\displaystyle{
-{\langle{\,\omega\,\,|\tilde H_w|D^{*0}}\rangle \over f_\pi}
\Biggl({m_{D}^2 - m_\omega^2 
\over m_{D^*}^2 - m_{\omega}^2}\Biggr)\sqrt{1\over 2}h_D}
$}\vspace{2mm}
\\
{$\displaystyle{
- {\langle{\hat\pi_H^+|\tilde H_w|D^+}\rangle \over f_\pi}
\Biggl({m_{D}^2 - m_\omega^2 
\over m_{D}^2 - m_{\hat\pi_H}^2}\Biggr)\sqrt{2}\hat g_H 
}$
}
\end{tabular}
\vspace{-3.5mm}
\\
 & &\\
\hline
\end{tabular}

\end{center}
%%%%%%%%%%%%%%%%%%%%%%%%%%%%%%%%%%%%%%%%%%%%%%%%%%%%%%%%%%%%%%%%%%%
\vspace{5mm}
%%%%%%%%%%%%%%%%%%%%%%%%%%%%%%%%%%%%%%%%%%%%%%%%%%%%%%%%%%%%%%%%%%%
is an approximate spectator decay, although four-quark states 
can contribute to its $s$-channel intermediate states. Since the spin 
and parity of the intermediate four quark state in this decay should 
be $J^{P} = 0^{-}$ and such a four-quark meson state includes some 
orbital excitation in the system, its mass and width will be much 
higher and broader than the corresponding $\{qq\bar q\bar q\}$ mesons 
with positive parity which have played an important role in two body 
decays of charm mesons~\cite{TBD99}. Therefore, pole contributions of 
four-quark mesons to the $D_s^+\rightarrow\phi\pi^+$ are not very 
important and the factorized amplitude for the decay can be estimated 
by using its measured rate.  For the same reason, we neglect possible 
four-quark meson contributions to a color mismatched decay, 
$D^+\rightarrow\phi\pi^+$, and will estimate its $D^*$ pole amplitude 
in the $u$-channel using its measured rate. For contributions of 
excited meson poles in the $u$-channel, however, we expect that they 
are small as discussed in Ref.~\cite{TBD99}. In the 
$D_s^+\rightarrow\omega\pi^+$, which is an approximate annihilation 
decay, its amplitude will be dominated by an $s$-channel pole 
contribution of hybrid meson ($\hat\pi_H$) with 
$I^GJ^{P(C)}=1^+0^{-(-)}$. 

As seen in Eq.~(\ref{eq:SURF}), the non-factorizable amplitudes are 
described in terms of {\it asymptotic matrix elements} (matrix 
elements taken between single hadron states with infinite momentum) 
of $A_P$ and $\tilde H_w$ since $M_{\rm ETC}$ has been discarded. 
Asymptotic matrix elements of $A_P$ are parameterized as 
%%%%%%%%%%%%%%%%%%%%%%%%%%%%%%%%%%%%%%%%%%%%%%%%%%%%%%%%%%%%%%%
$\sqrt{2}\langle{D^{*0}|A_{\pi^-}|D^+}\rangle=-h_D$ and 
$\langle{\omega|A_{\pi^-}|\hat\pi_H^+}\rangle=\sqrt{2}\hat g_H$. 
%%%%%%%%%%%%%%%%%%%%%%%%%%%%%%%%%%%%%%%%%%%%%%%%%%%%%%%%%%%%%%%%
However, our result on the $D^+\rightarrow\omega\pi^+$ in this short 
note does not explicitly depend on these matrix elements. 

Constraints on (i.e., selection rules of) asymptotic matrix elements 
of $\tilde H_w$ can be obtained by using an intuitive quark 
counting~\cite{TBD99} which is an analogue to the 
Miura-Minamikawa-Pati-Woo theorem in hyperon decays~\cite{MMPW}. 
Noting that the wave function of the ground-state $\{q\bar q\}_0$ 
meson is anti-symmetric under the exchange of its constituent quark and 
anti-quark~\cite{CLOSE}, we obtain~\cite{TBD99} 
%%%%%%%%%%%%%%%%%%%%%%%%%%%%%%%%%%%%%%%%%%%%%%%%%%%%%%%%%%%%%%%%%
$\langle{\{q\bar q\}_0|\tilde O_+|\{q\bar q\}_0}\rangle = 0$, 
%%%%%%%%%%%%%%%%%%%%%%%%%%%%%%%%%%%%%%%%%%%%%%%%%%%%%%%%%%%%%%%%%
which leads to 
%%%%%%%%%%%%%%%%%%%%%%%%%%%%%%%%%%%%%%%%%%%%%%%%%%%%%%%%%%%%%%%%%%%%
\begin{eqnarray}
&& \langle{\pi^+|\tilde H_w|D_s^+}\rangle 
= -\langle{\bar K^0|\tilde H_w|D^0}\rangle,             \nonumber\\
&& \langle{\rho^+|\tilde H_w|D_s^{*+}}\rangle 
= - \langle{\bar K^{*0}|\tilde H_w|D^{*0}}\rangle,
                                                         \nonumber\\
&&  \langle{\pi^+|\tilde H_w|D^+}\rangle 
= \sqrt{2}\langle{\pi^0|\tilde H_w|D^0}\rangle
= -\langle{K^{+}|\tilde H_w|D_s^{+}}\rangle =\cdots , 
                                                       \nonumber\\
&& \langle{\rho^+|\tilde H_w|D^{*+}}\rangle 
= \sqrt{2}\langle{\rho^0|\tilde H_w|D^{*0}}\rangle
= -\langle{K^{*+}|\tilde H_w|D_s^{*+}}\rangle  \nonumber\\
&& \hspace{25mm}
= -\sqrt{2}\langle{\omega|\tilde H_w|D^{*0}}\rangle   
= \langle{\phi|\tilde H_w|D^{*0}}\rangle = \cdots , 
                                                     \label{eq:SR}
\end{eqnarray}
%%%%%%%%%%%%%%%%%%%%%%%%%%%%%%%%%%%%%%%%%%%%%%%%%%%%%%%%%%%%%%%%%%%%
where approximate equalities $V_{ud}=V_{cs}$ and $V_{us}=-V_{cd}$ 
have been used. Asymptotic flavor $SU_f(3)$ symmetry~\cite{ASYMP} 
relates the asymptotic matrix elements of the CKM-angle favored weak 
Hamiltonian to those of the CKM-angle suppressed one~\cite{TBD99}, 
for example, 
%%%%%%%%%%%%%%%%%%%%%%%%%%%%%%%%%%%%%%%%%%%%%%%%%%%%%%%%%%%%%%%%%%%%
\begin{equation}
\langle{\pi^+|\tilde H_w|D^+}\rangle
= {V_{cd} \over V_{cs}}\langle{\pi^+|\tilde H_w|D_s^+}\rangle,\,\, 
{\rm etc.}                    \label{eq:SU(3)}
\end{equation}
%%%%%%%%%%%%%%%%%%%%%%%%%%%%%%%%%%%%%%%%%%%%%%%%%%%%%%%%%%%%%%%%%%%%%

Inserting the above parameterizations of asymptotic matrix elements 
of $\tilde H_w$ and $A_P$'s into the non-factorizable hard PS meson 
amplitudes, we obtain amplitudes for the annihilation and color 
mismatched types of $D\rightarrow VP$ decays whose surface terms are 
dominated by pole contributions of the ground-state $\{q\bar q\}_0$ 
and hybrid $\{q\bar qg\}$ mesons. In the approximation in which pole 
contributions of $\{q\bar q\}_0$ and hybrid $J^{P(C)}=0^{-(-)}$ 
mesons are taken into account, the non-factorizable amplitudes for
typical quasi two body decays are given in the third column of 
Table~I, where the useless imaginary unit has been factored out, 
i.e., $M_{\rm NF} = i{\cal M}_{\rm NF}$. The amplitudes for the 
$D_s^+\rightarrow\rho^0\pi^+$ and $D_s^+\rightarrow\omega\pi^+$ are 
described by two annihilation diagrams and they cancel each other in 
the former while they interfere constructively with each other in 
the latter~\cite{Chau}. 

To estimate the rate for the $D^+\rightarrow\omega\pi^+$ decay, we
take the central values of the measured ones~\cite{PDG00} of the CKM 
matrix elements [$V_{ud}$, $V_{us}$, $V_{cd}$, $V_{cs}$], decay 
constants [$f_\pi$, $f_K$], and lifetimes of charm mesons 
[$\tau(D^+)$, $\tau(D^0)$, $\tau(D_s^+)$]. As seen in Table~I, 
%%%%%%%%%%%%%%%%%%%%%%%%%%%%%%%%%%%%%%%%%%%%%%%%%%%%%%%%%%%%%%%%%%%%
$M(D^+\rightarrow\phi\pi^+)
\simeq M^{(D^*)}(D^+\rightarrow\phi\pi^+)$.  
%%%%%%%%%%%%%%%%%%%%%%%%%%%%%%%%%%%%%%%%%%%%%%%%%%%%%%%%%%%%%%%%%%%
Eliminating useless imaginary unit, we obtain 
%%%%%%%%%%%%%%%%%%%%%%%%%%%%%%%%%%%%%%%%%%%%%%%%%%%%%%%%%%%%%%%%%%%%%
\begin{eqnarray}
&&{\cal M}^{(D^*)}(D^+\rightarrow\omega\pi^+)     
\simeq 1.03\Biggl[{\langle{\omega|\tilde H_w|D^{*0}}\rangle
         \over \langle{\phi|\tilde H_w|D^{*0}}\rangle}\Biggr]
{\cal M}(D^+\rightarrow\phi\pi^+) 
\simeq \bigl\{-5.3\bigr\}\times 10^{-7}\,\,{\rm GeV}
\end{eqnarray}
%%%%%%%%%%%%%%%%%%%%%%%%%%%%%%%%%%%%%%%%%%%%%%%%%%%%%%%%%%%%%%%%%%%%%
using Eq.~(\ref{eq:SR}) and 
%%%%%%%%%%%%%%%%%%%%%%%%%%%%%%%%%%%%%%%%%%%%%%%%%%%%%%%%%%%%%%%%%%%
$|{\cal M}(D^+\rightarrow\phi\pi^+)|\simeq 7.4 \times10^{-7}$ GeV 
%%%%%%%%%%%%%%%%%%%%%%%%%%%%%%%%%%%%%%%%%%%%%%%%%%%%%%%%%%%%%%%%%%%%%
from 
%%%%%%%%%%%%%%%%%%%%%%%%%%%%%%%%%%%%%%%%%%%%%%%%%%%%%%%%%%%%%%%%%%%%
$\Gamma(D^+\rightarrow\phi\pi^+)_{\rm exp}
=(4.0 \pm 0.4)\times 10^{-15}$ GeV~\cite{PDG00}, 
%%%%%%%%%%%%%%%%%%%%%%%%%%%%%%%%%%%%%%%%%%%%%%%%%%%%%%%%%%%%%%%%%%%%
where the sign of $\langle{\rho^+|\tilde H_w|D^{*+}}\rangle$ [and
hence $\langle{\phi|\tilde H_w|D^{*0}}\rangle$ because of 
Eq.~(\ref{eq:SR})] have been taken to be the same as that of 
$\langle{\pi^+|\tilde H_w|D^+}\rangle$ in Ref.~\cite{TBD99} as was in 
weak interactions of $K$ mesons~\cite{Dalitz}, i.e., 
%%%%%%%%%%%%%%%%%%%%%%%%%%%%%%%%%%%%%%%%%%%%%%%%%%%%%%%%%%%%%%%%%%%%%
$\langle{\rho^0|\tilde H_w|K^{*0}}\rangle
/\langle{\pi^0|\tilde H_w|K^0}\rangle > 0$. 
%%%%%%%%%%%%%%%%%%%%%%%%%%%%%%%%%%%%%%%%%%%%%%%%%%%%%%%%%%%%%%%%%%%%%%

The amplitude for the $D_s^+\rightarrow\omega\pi^+$ decay is
dominated by a pole contribution of $\hat\pi_H$, i.e., 
%%%%%%%%%%%%%%%%%%%%%%%%%%%%%%%%%%%%%%%%%%%%%%%%%%%%%%%%%%%%%%%%%%%%
${\cal M}(D_s^+\rightarrow\omega\pi^+) \simeq 
{\cal M}^{(\hat\pi_H)}(D_s^+\rightarrow\omega\pi^+)$. 
%%%%%%%%%%%%%%%%%%%%%%%%%%%%%%%%%%%%%%%%%%%%%%%%%%%%%%%%%%%%%%%%%%%
From the measured rate~\cite{PDG00}, 
%%%%%%%%%%%%%%%%%%%%%%%%%%%%%%%%%%%%%%%%%%%%%%%%%%%%%%%%%%%%%%%%%%%%
$\Gamma(D_s^+\rightarrow\omega\pi^+)_{\rm exp}
=(3.9 \pm 1.5)\times 10^{-15}$ GeV, 
%%%%%%%%%%%%%%%%%%%%%%%%%%%%%%%%%%%%%%%%%%%%%%%%%%%%%%%%%%%%%%%%%%%%
we obtain 
%%%%%%%%%%%%%%%%%%%%%%%%%%%%%%%%%%%%%%%%%%%%%%%%%%%%%%%%%%%%%%%%
$|{\cal M}^{(\hat\pi_H)}(D_s^+\rightarrow\omega\pi^+)|
\simeq 6.8\times 10^{-7}$ GeV.  
%%%%%%%%%%%%%%%%%%%%%%%%%%%%%%%%%%%%%%%%%%%%%%%%%%%%%%%%%%%%%%%%%%%
Using the asymptotic $SU_f(3)$ relation in Eq.(\ref{eq:SU(3)}), we 
obtain 
%%%%%%%%%%%%%%%%%%%%%%%%%%%%%%%%%%%%%%%%%%%%%%%%%%%%%%%%%%%%%%%%%%
\begin{equation}
{\cal M}^{(\hat\pi_H)}(D^+\rightarrow\omega\pi^+)
\simeq {-1.3\times \Bigl(
{m_{D_s}^2 - m_{\hat\pi_H}^2 \over m_D^2 - m_{\hat\pi_H}^2}
\Bigr)
\hat\alpha_H\times 10^{-7}}\,\,{\rm GeV}
                                              \label{eq:pi_H-pole}
\end{equation}
%%%%%%%%%%%%%%%%%%%%%%%%%%%%%%%%%%%%%%%%%%%%%%%%%%%%%%%%%%%%%%%%%%%
where $\hat\alpha_H$ is a parameter providing the sign of 
${\cal M}(D_s^+\rightarrow\omega\pi^+)$. 

The factorized amplitude for the $D^+\rightarrow\omega\pi^+$ includes
a form factor $A_0^{(\omega D)}(m_\pi^2)$. To estimate it, we put 
%%%%%%%%%%%%%%%%%%%%%%%%%%%%%%%%%%%%%%%%%%%%%%%%%%%%%%%%%%%%%%%%%%%
$A_0^{(\omega D)}(m_\pi^2) \simeq A_0^{(\phi D_s)}(m_\pi^2)$  
%%%%%%%%%%%%%%%%%%%%%%%%%%%%%%%%%%%%%%%%%%%%%%%%%%%%%%%%%%%%%%%%%%
which can be obtained by applying (asymptotic) 
%%%%%%%%%%%%%%%%%%%%%%%%%%%%%%%%%%%%%%%%%%%%%%%%%%%%%%%%%%%%%%%%%%%%%%
\newpage
\begin{center}
\hspace{-0mm}
\epsfysize = 2.2in
\hspace{-0mm}
\epsfxsize = 2.7in
\epsfysize = 2.5in
\epsffile{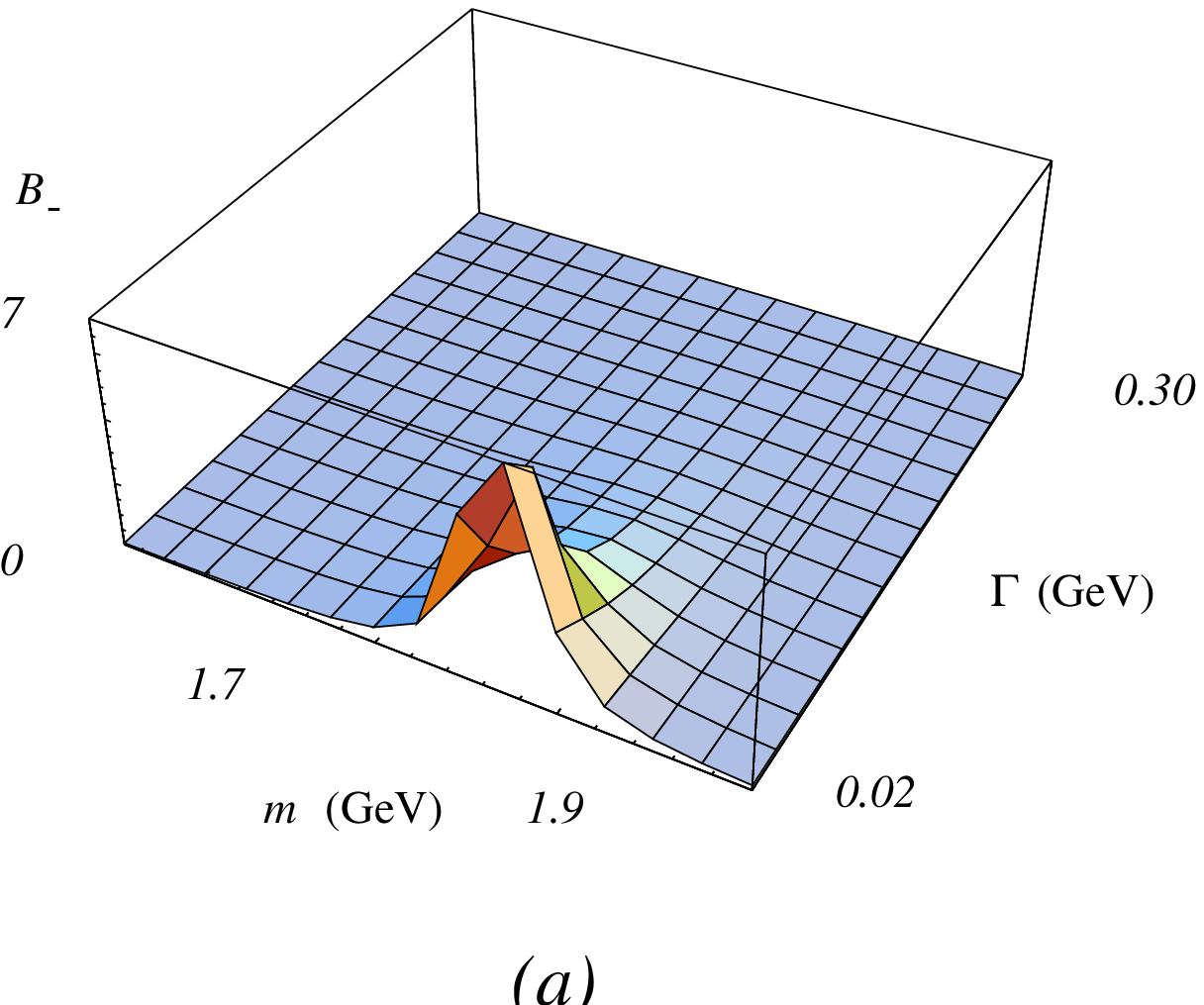}
\hspace{10mm}
\epsfxsize = 2.7in
\epsfysize = 2.5in
\epsffile{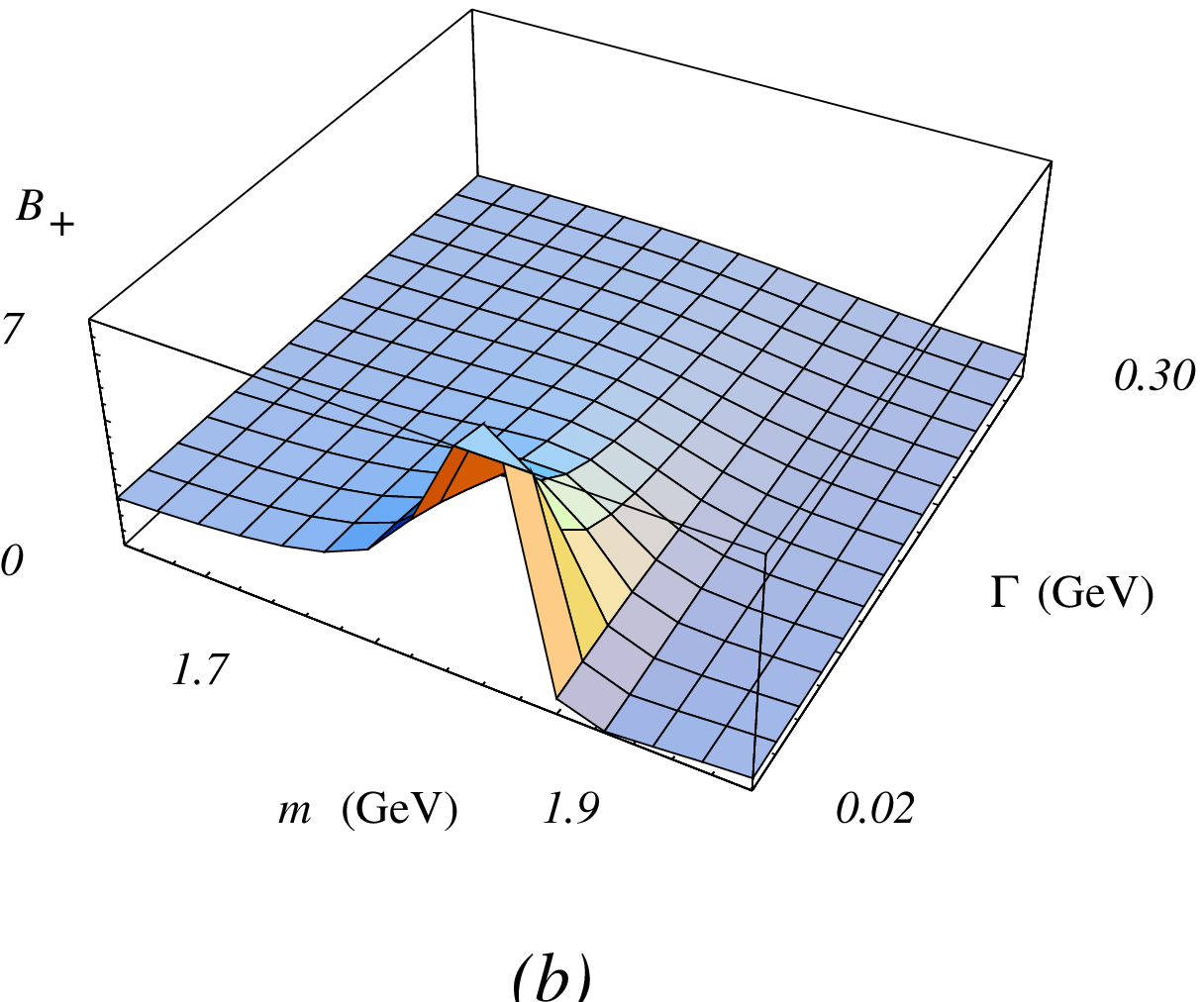}
\vspace{5mm}
\begin{minipage}{140mm}
{Fig.~I. ${B}(D^+\rightarrow\omega\pi^+)\times 10^{3}$ vs. 
($m=m_{\hat\pi_H},\Gamma=\Gamma_{\hat\pi_H})$. ({\it a}) $B_-$ for 
$\hat\alpha_H=-1$ and ({\it b}) $B_+$ for $\hat\alpha_H=+1$.}
\end{minipage}
\end{center}
\vspace{5mm}
%%%%%%%%%%%%%%%%%%%%%%%%%%%%%%%%%%%%%%%%%%%%%%%%%%%%%%%%%%%%%%%%%%%%%
$SU_f(3)$ symmetry
to the current matrix elements. Then, comparing our amplitude for 
the $D_s^+\rightarrow\phi\pi^+$ decay with 
%%%%%%%%%%%%%%%%%%%%%%%%%%%%%%%%%%%%%%%%%%%%%%%%%%%%%%%%%%%%%%%%%%
$\Gamma(D_s^+\rightarrow\phi\pi^+)_{\rm exp}
=(5.0\pm 1.3)\times 10^{-14}$ GeV, 
%%%%%%%%%%%%%%%%%%%%%%%%%%%%%%%%%%%%%%%%%%%%%%%%%%%%%%%%%%%%%%%%%%
we obtain $|A_0^{(\phi D_s)}(m_\pi^2)| \simeq 0.76$. It is reasonable 
since it is not very far from two predicted values~\cite{Kamal}, 
$A_0^{(\phi D_s)}(0)\simeq 0.70$ and $\simeq 0.85$, by two different 
models, BSW~\cite{BSW} and ISGW~\cite{ISGW}, respectively. In this 
way, we obtain approximately 
%%%%%%%%%%%%%%%%%%%%%%%%%%%%%%%%%%%%%%%%%%%%%%%%%%%%%%%%%%%%%%%%%%%%%
\begin{equation}
{\cal M}_{\rm tot}(D^+\rightarrow\omega\pi^+)
\simeq  \Bigl\{{3.6}a_1 
           - {5.3} - 1.3\hat\alpha_H
                         \Bigl({m_{D_s}^2 - m_{\hat\pi_H}^2 
                               \over m_D^2 - m_{\hat\pi_H}^2}\Bigr)
\Bigr\}\times 10^{-7}\,\,{\rm GeV},          \label{eq:M_tot}
\end{equation}
%%%%%%%%%%%%%%%%%%%%%%%%%%%%%%%%%%%%%%%%%%%%%%%%%%%%%%%%%%%%%%%%%%%%%%%
where the first, second and third terms on the right-hand-side of 
Eq.(\ref{eq:M_tot}) are from ${\cal M}_{\rm FA}$, ${\cal M}^{(D^*)}$ 
and ${\cal M}^{(\hat\pi_H)}$, respectively. Taking 
$a_1=1.09$~\cite{Buras}, we will study the branching ratio for the 
$D^+\rightarrow\omega\pi^+$. If we neglected the non-factorizable 
contributions, we would obtain $B(D^+\rightarrow\omega\pi^+)_{\rm FA}
\simeq 2.1\times 10^{-3}$. When we take account for them, however, 
we predict $B(D^+\rightarrow\omega\pi^+)$ as a function of mass 
$m_{\hat\pi_H}$ and width $\Gamma_{\hat\pi_H}$ of $\hat\pi_H$. 
It depends on the relative sign between $M_{\rm FA}$ and 
$M^{(\hat\pi_H)}$. As seen in Fig.~I, however, gross features of the 
branching ratios, $B_-$ for $\hat\alpha_H = -1$ and $B_+$ for 
$\hat\alpha_H = +1$, are not very different from each other since 
${\cal M}^{(D^*)}$ cancel almost ${\cal M}_{\rm FA}$ in the present 
case. Except for the neighborhood of the singular point, 
%%%%%%%%%%%%%%%%%%%%%%%%%%%%%%%%%%%%%%%%%%%%%%%%%%%%%%%%%%%%%%%%%%%%
$(m_{\hat\pi_H}=m_D,\,\,\Gamma_{\hat\pi_H}=0)$, 
%%%%%%%%%%%%%%%%%%%%%%%%%%%%%%%%%%%%%%%%%%%%%%%%%%%%%%%%%%%%%%%%%%%%
both of $B_-$ and $B_+$ are always lower than the experimental upper 
bound [${B}(D^+\rightarrow\omega\pi^+)_{\rm exp}< 7\times 10^{-3}$] 
and are very sensitive to the values of $m_{\hat\pi_H}$ and 
$\Gamma_{\hat \pi_H}$ in the region, 
%%%%%%%%%%%%%%%%%%%%%%%%%%%%%%%%%%%%%%%%%%%%%%%%%%%%%%%%%%%%%%%%%%%%%
($1.8\lesssim m_{\hat\pi_H}\lesssim 1.9$ GeV,  
$\Gamma_{\hat \pi_H}\lesssim 0.10$ GeV), and 
($1.75\lesssim m_{\hat\pi_H}\lesssim 1.9$ GeV,  
$\Gamma_{\hat \pi_H}\lesssim 0.15$ GeV), 
%%%%%%%%%%%%%%%%%%%%%%%%%%%%%%%%%%%%%%%%%%%%%%%%%%%%%%%%%%%%%%%%%%%%%%
respectively. Outside of the above regions, however, they are very
small, 
i.e., ${B}_{+}$ is around the level of 
$\sim 1.5\times 10^{-3}$ for $m_{\hat\pi_H}\lesssim 1.8$ GeV and much 
smaller for $m_{\hat\pi_H}\gtrsim 1.9$ GeV while 
${B}_{-}\lesssim 0.5\times 10^{-3}$ outside the region, 
($1.8\lesssim m_{\hat\pi_H}\lesssim 1.9$ GeV, 
$\Gamma_{\hat \pi_H}\lesssim 0.10$ GeV). 
%%%%%%%%%%%%%%%%%%%%%%%%%%%%%%%%%%%%%%%%%%%%%%%%%%%%%%%%%%%%%%%%%%%%

In summary, the $D^+\rightarrow\omega\pi^+$ decay has been studied 
by decomposing the amplitude into a sum of factorizable and 
non-factorizable ones. The former has been estimated by using the 
naive factorization in the BSW scheme while the latter has been 
calculated by using a hard pion approximation. As the result, its 
amplitude has been approximately given by a sum of $M_{\rm FA}$, 
$M^{(D^*)}$ and $M^{(\hat\pi_H)}$. The first two have been estimated 
by using the measured rates for the $D_s^+ \rightarrow \phi\pi^+$ and 
$D^+ \rightarrow \phi\pi^+$. We have predicted 
$B(D^+\rightarrow\omega\pi^+)$ as a function of mass and width of a 
hypothetical hybrid meson $\hat\pi_H$ with 
$I^GJ^{P(C)} = 1^+0^{-(-)}$. If $m_{\hat\pi_H}$ is close to $m_D$ and 
$\Gamma_{\hat\pi_H}$ is small, it will be strongly enhanced. If not, 
however, it will be $\sim 1.5\times 10^{-3}$ or much smaller. 
The former is not very far from the result by the naive
factorization. 
%%%%%%%%%%%%%%%%%%%%%%%%%%%%%%%%%%%%%%%%%%%%%%%%%%%%%%%%%%%%%%%%%%%%%
\vspace{5mm}

\thanks{
The author would like to thank Prof. T.~E.~Browder for attracting
his attention to the $D^+ \rightarrow \omega\pi^+$ decay. He also
would like to appreciate Prof. H. Yamamoto for discussions and 
comments. 
}

%%%%%%%%%%%%%%%%%%%%%%%%%%%%%%%%%%%%%%%%%%%%%%%%%%%%%%%%%%%%%%%%%%


\begin{references}
\bibitem{ASYMP} A comprehensive review on the asymptotic flavor 
symmetry and its fruitful results has been provided in
Ref.~\cite{suppl}.
\end{references}
\end{document}